\documentclass[12pt]{article}
\usepackage{a4wide}
\usepackage{amssymb}
\usepackage{graphicx}
\begin{document}
{\renewcommand{\thefootnote}{\fnsymbol{footnote}}
\medskip
\begin{center}
{\LARGE Factor ordering and large-volume
  dynamics\\ in quantum cosmology}\\
\vspace{1.5em}
Martin Bojowald\footnote{e-mail address: {\tt bojowald@gravity.psu.edu}}
and David Simpson\footnote{e-mail address: {\tt tertsu@gmail.com}}
\\
\vspace{0.5em}
Institute for Gravitation and the Cosmos,
The Pennsylvania State
University,\\
104 Davey Lab, University Park, PA 16802, USA\\
\vspace{1.5em}
\end{center}
}

\setcounter{footnote}{0}

\begin{abstract}
  Quantum cosmology implies corrections to the classical equations of motion
  which may lead to significant departures from the classical trajectory,
  especially at high curvature near the big-bang singularity. Corrections
  could in principle be significant even in certain low-curvature regimes,
  provided that they add up during long cosmic evolution.  The analysis of
  such terms is therefore an important problem to make sure that the theory
  shows acceptable semiclassical behavior. This paper presents a general
  search for terms of this type as corrections in effective equations for a
  $k=0$ isotropic quantum cosmological model with a free, massless scalar
  field. Specifically, the question of whether such models can show a collapse
  by quantum effects is studied, and it turns out that factor-ordering choices
  in the Hamiltonian constraint are especially relevant in this regard.  A
  systematic analysis of factor-ordering ambiguities in effective equations is
  therefore developed.
\end{abstract}

\section{Introduction}

Effective equations are useful tools to analyze quantum theories by adding
suitable correction terms to classical equations, for instance by using
equations of motion for expectation values of operators of interest. Just like
expectation values, effective equations that describe their evolution depend
on the states or class of states whose evolution is being approximated. The
low-energy effective action often used in particle physics, as perhaps the
best-known example, describes quantum corrections for states near the vacuum
of the interacting theory. For other states, different quantum corrections
arise. Effective equations are state-independent only in rare cases such as
free theories or the harmonic oscillator, when no quantum back-reaction and
therefore no dynamical quantum corrections occur. Such systems also exist in
(loop) quantum cosmology, where they play a similar role as the base order for
perturbation theories when interactions are present \cite{BouncePert}.

As an example for potentially large perturbative effects in quantum cosmology,
it has been suggested that classically ever-expanding models could eventually
collapse due to quantum corrections \cite{EffRecollapse}. Such a feature is
unexpected because the more the universe expands, the more classical it is
supposed to become, with quantum corrections playing smaller and smaller
roles. However, with long evolution involved, small quantum effects could
potentially add up and eventually drive the universe into collapse even in
semiclassical regimes. As shown in \cite{EffRecollapse}, the outcome
depends on what semiclassical quantum state is realized and how its
fluctuations of different variables change in time. To test the proposed
scenario, one needs information about dynamical semiclassical states over long
evolution times, an issue that requires good control on quantum evolution and,
when addressed with effective equations, a systematic scheme of going beyond
the leading classical order.

The effective equations used in \cite{EffRecollapse} (as well as
\cite{Taveras}) were an extension of the classical limit obtained from methods
provided in \cite{Bohr}. This scheme of going beyond classical order, however,
is not entirely systematic because assumptions about the evolving state must
be made; they have not been self-consistently derived within the scheme. For
this reason, the suggested collapse remained a possibility but could not be
demonstrated firmly. (Note that going to the other extreme, the high densities
of the big bang, requires even better control of evolving states no longer
required to be semiclassical. For this reason, the high-curvature regime of
quantum cosmology remains poorly understood, even though rather
definite-looking statements are sometimes attempted.) In this article, we use
a systematic perturbative framework to derive information about effective
equations and dynamical semiclassical states and see whether the suggestions
of \cite{EffRecollapse} can be realized. We will be led to a more systematic
look on factor-ordering ambiguities than provided so far.

\section{Modified Friedmann equation, recollapse, and factor ordering}

Although the low-curvature regime can be formulated in Wheeler--DeWitt quantum
cosmology \cite{QCReview} and does not require additional corrections and
modifications from extensions such as loop quantum cosmology
\cite{LivRev,Springer}, we will follow \cite{EffRecollapse} and use the latter
for a general discussion. This choice also helps us keep our notation close to
that of \cite{EffRecollapse}.  In brief terms, loop quantum cosmology of
spatially flat isotropic models proceeds by formulating the Friedmann equation
in canonical variables\footnote{We assume a fixed coordinate volume $V_0$ of
  the homogeneous region considered for a minisuperspace model, normalized to
  $V_0=1$. Classically, $V_0$ can be changed (non-canonically) without
  affecting the dynamics, but this feature is violated after quantization,
  where no unitary transformation exists to change $V_0$ \cite{Springer}. This
  issue is a minisuperspace limitation which we can ignore in this
  paper because we will be interested only in qualitative effects. Note that
  no minisuperspace quantization can overcome this limitation, although,
  compared to loop quantum cosmology, it is less severe in Wheeler--DeWitt
  models with continuous minisuperspace.}  $c=\gamma\dot{a}$ and $|p|=a^2$,
\begin{equation}
\label{Friedmann}
- \frac {3}{8\pi G \gamma^2} c^2 \sqrt{ |p| } + H_{\rm matter} = 0\,,
\end{equation}
and modifying it by using periodic functions of $c$ (or, as seen in more
detail below, $\delta(p)c$ with some function $\delta(p)$) instead of $c$. In
this way, one mimics the fact that the full theory of loop quantum gravity
\cite{Rov,ThomasRev,ALRev} provides operators only for holonomies of the
connection corresponding to $c$, not for the SU(2)-connection itself
\cite{LoopRep}.

The (almost) periodic nature of functions substituting $c$ is motivated by a
reduction of the internal gauge freedom from SU(2) to U(1) when isotropy is
imposed \cite{SymmRed}. Recent attempts to bring about a closer relation
between minisuperspace models of loop quantum cosmology and the full theory
have questioned the almost periodic nature of isotropic holonomies as a
reliable feature of loop quantum gravity \cite{NonAb}, as a consequence of
crucial differences between Abelian and non-Abelian treatments
\cite{DegFull}. In this paper, we aim to investigate a question that can be
posed within pure minisuperspace models, whose form and quantum modifications
we will take for granted. We are not claiming that the features analyzed are
in any way related to the full theory of loop quantum gravity.

Such a modification to
\begin{equation}
\label{modconstraint}
- \frac {3}{8\pi G \gamma^2}  \frac{\sin(\delta(p)c)^2}{\delta(p)^2}
 \sqrt{|p|}  + H_{\rm matter} = 0
\end{equation}
is most relevant in regimes where $\delta(p)c$ is large (at large
curvature). If one expands the sine function, terms beyond the classical ones
may be interpreted as contributions to higher-curvature terms. (They are not
complete as higher-curvature corrections, however, because higher time
derivatives have not been included. The latter would result from quantum
back-reaction \cite{Karpacz,HigherTime}.) From equations of motion generated
by (\ref{modconstraint}) used as a Hamiltonian constraint in proper-time
gauge, one finds that the modified Friedmann equation in terms of $\dot{a}$ or
$H=\dot{a}/a$ takes the form
\begin{equation}
H^2 = \frac{8 \pi G}{3} \rho \left( 1 - \frac{
    \rho}{\rho_\mathrm{QG}}\right)
\end{equation}
with $\rho_{\rm QG}=8\pi G/3\delta^2|p|$
\cite{AmbigConstr,RSLoopDual,APSII,BounceSqueezed}. If one assumes
$\delta(p)\propto 1/\sqrt{|p|}$ as in \cite{APSII}, $\rho_{\rm QG}$ is a
constant and Planckian for $\sqrt{|p|}\delta(p)\sim \ell_{\rm P}$. This form
makes it especially clear that the modification is relevant only at large
density.

Loop quantum cosmology represents the modified constraint
(\ref{modconstraint}) as a difference equation \cite{cosmoIV,IsoCosmo},
obtained from shift operators $\widehat{\exp(i\delta(p)c)}$ acting on a wave
function in the triad representation. From this difference equation or the
underlying Hamiltonian-constraint operator, (\ref{modconstraint}) can be
reproduced as the ``classical'' limit by computing the expectation value in
Gaussian states and ignoring all fluctuation-dependent terms \cite{Bohr}. From
the point of view of quantum dynamics, the result is the classical limit
because fluctuations and their back-reaction on expectation values are
ignored. Still, the result does not agree with the classical equation
(\ref{Friedmann}) because loop quantum gravity implies quantum-geometry
corrections in addition to the usual dynamical quantum corrections, which give
rise to the modifiction (\ref{modconstraint}) of (\ref{Friedmann}). Only a
combined classical and low-curvature limit \cite{SemiClass} brings loop
quantum cosmology fully back to the classical Friedmann equation
(\ref{Friedmann}). (See also \cite{RegularizationFRW,RegularizationLQC}.)

The modification in (\ref{modconstraint}) alone does not affect late-time
evolution much. However, quantum evolution implies additional corrections,
which in canonical models can be computed by quantum back-reaction of
fluctuations and higher moments of a state on expectation values
\cite{EffAc,Karpacz}. The leading-order expectation value that gives rise to
the classical limit must be expanded systematically to include these terms.
Such terms were computed in \cite{EffRecollapse}, with potentially significant
low-curvature implications.  In addition to the usual quantum corrections on
small distance scales, higher-order corrections may allow for the possibility
of an expanding universe to collapse at large size when the energy density is
suitably small, even in the absence of spatial curvature --- a significant
departure from classical theory. (The possibility of a recollapse was also
conjectured in \cite{AltCollapse} by an alternative quantum Hamitonian scheme,
as well as in \cite{FuncCollapse1,FuncCollapse2} by a coherent state
functional integral approach. In the latter cases, assumptions on the
asymptopic semiclassical behavior of coherent states were more restrictive.)

In \cite{EffRecollapse}, lacking a derivation of properties of dynamical
semiclassical states, a suitable coherent state for large volume and late
times had to be chosen, assumed, as in most cases, to be an uncorrelated
Gaussian. With this choice, the classical-limit scheme of \cite{Bohr} was
developed further to calculate fluctuations and expectation values. The end
result is a modified Friedmann equation, which we write here as
\begin{equation}
\label{ym friedmann}
H^2 = \frac{8 \pi G}{3} \rho \left[ 1 - \frac{ \rho}{\rho_\mathrm{QG}}
  \left( 1 +  \bar{\epsilon}^2 \right) + \frac{1}{2}\bar{\epsilon}^2
  - 2 \epsilon^2 \frac{\rho_\mathrm{QG}}{\rho} \right] \,.
\end{equation}
In this expression, $\epsilon$ is the quantum fluctuation of curvature or
the Hubble parameter $H$, while $\bar{\epsilon}:=(\Delta \nu)/\nu$ is the
relative volume fluctuation, using $\nu=a^3$. (Unlike $\bar{\epsilon}$, the
latter is not invariant under a change of spatial coordinates or of $V_0$.)
As mentioned, Eq. (\ref{ym friedmann}) was derived with a Gaussian ansatz for
the wave function with zero correlations of the canonical pair $(H,\nu)$, for
which $1/\epsilon$, the inverse curvature fluctuation, is proportional to the
volume fluctuation $\Delta\nu=\nu\bar{\epsilon}$ thanks to the uncertainty
relation.

The last term in (\ref{ym friedmann}) is especially important for the question
posed here, because it has the lowest power of the energy density $\rho$.  The
equation has the correct classical form when all fluctuations are sent to zero
first, and then $\rho_{\rm QG}$ to infinity.  The semiclassical behavior, in
which fluctuations are small but not zero, agrees with the classical one
provided $\epsilon$ goes sufficiently fast to zero as the energy density
decreases in an expanding universe. For a Gaussian state, $\Delta \nu$ would
increase accordingly, but in (\ref{ym friedmann}) it is always suppressed by
factors of $1/\nu$ in $\bar{\epsilon}$. Indeed, one can solve the quantum
model in a specific factor ordering and find properties of dynamical coherent
states \cite{BounceCohStates}. In this harmonic quantum model, with the same
classical dynamics, $\epsilon$ must decrease at low curvature and cannot
remain constant, while $\bar{\epsilon}$ is asymptotically constant for any
state \cite{BouncePert}. This behavior may, however, change when the model is
quantized with a different choice of factor ordering.  The question is whether
it is then possible for $\epsilon$ to decrease sufficiently slowly that the
last term in Eq. (\ref{ym friedmann}) can cancel the others and enable a
recollapse, where $H=0$ at large $\nu$.

In the present context, the important part to note from Eq.~(\ref{ym
  friedmann}) is that at late time and large volume, for an expanding universe
with suitably low energy density ($\rho \sim 2 \epsilon^2 \rho_\mathrm{QG}$),
the universe could perhaps collapse. An inverse-$\rho$ quantum correction
associated with $\epsilon$ in (\ref{ym friedmann}) may seem surprising. But
one way of looking at this is by distributing the $\rho$ in front to see that
the final term is independent of the energy density and only depends on the
quantum fluctuations. It functions as a negative cosmological-constant term if
$\epsilon$ varies slowly. When the energy density is of the order of those
fluctuations, the final term dominates and could lead to collapse -- depending
on the exact quantum dynamics.

From the perspective of effective equations, the classical model quantized in
\cite{EffRecollapse} --- a spatially flat isotropic model sourced by a free,
massless scalar --- is identical to a harmonic one, without any quantum
back-reaction \cite{BouncePert}. The only corrections to classical evolution
are due to quantum geometry; no quantum back-reaction should occur. However,
compared to \cite{BouncePert}, \cite{EffRecollapse} used a different factor
ordering of the quantum constraint as well as quantum-geometry modifications
of inverse-triad type, not just holonomy corrections as in
(\ref{modconstraint}). These variations imply quantum back-reaction, which,
thanks to the proximity to a harmonic model, should not be large but could
still be significant after long evolution. In this paper, we explore the
possibility of such a term in the modified Friedmann equation found by
embarking on a systematic method of deriving effective equations, including
the underlying properties of dynamical semiclassical states
\cite{EffAc,Karpacz}. By way of the example of collapse solutions, we will
therefore study several implications of terms in effective equations,
including the role of factor ordering. We will provide a general
parameterization of these ambiguities in an expansion of the quantum-corrected
Hamiltonian constraint by powers of $H$, which turn out to require inverse
negative powers as well. These methods will require a shift in viewpoint,
focusing on the form of the last term in (\ref{ym friedmann}) instead of the
possibility of a recollapse which is difficult to assess in an expansion with
negative powers of $H$. This fact highlights how subtle the semiclassical
limit of quantum cosmology can be. In particular, it is not clear whether
(loop) quantum cosmology has the standard semiclassical behavior even in
isotropic models.

\section{Quantum cosmology and effective equations}

In spatially flat isotropic models, the basic pair of canonical variables is
the extrinsic curvature $c = \gamma \, {\rm d} a / {\rm d} \tau $ and its
conjugate $|p| = a^2$ where $\gamma$ is the Barbero-Immirzi parameter
\cite{AshVarReell,Immirzi} of loop quantum gravity, $a$ is the scale
factor, and $\tau$ is proper time. The momentum $p$ is derived from an
isotropic densitized triad and so can take both signs due to the triad
orientation. The fundamental Poisson bracket is $ \{ c, p \} = 8 \pi G \gamma
/ 3$ where $G$ is Newton's constant as before. (As already mentioned, we
assume the coordinate volume of our homogeneous region to equal one.)
Beginning with the Friedmann equation for an isotropic and homogeneous
background, we can write the Hamiltonian constraint in these canonical
variables as (\ref{Friedmann}) with $H_{\rm matter} = \frac{1}{2} |p|^{-3/2}
p_{\phi}^2$ for a free, massless scalar $\phi$ where $p_\phi$ is the momentum
conjugate to $\phi$ such that $ \{ \phi, p_\phi \} = 1$.

\subsection{The model}

As this work is motivated by \cite{EffRecollapse}, we will use a similar
notation for a direct comparison, and so by a canonical transformation we
introduce new conjugate variables:
\begin{equation}
\label{b and v}
b \equiv \delta \frac{ c} { \sqrt{ |p| } } \quad , \quad 
    \nu \equiv \frac{2} { 3 \delta} {\rm sgn}( p ) |p|^{3/2} 
\quad \mbox{such that} \,   \{ b , \nu \} = 8 \pi G \gamma / 3\,.
\end{equation}
Here, $\delta $ is a length parameter that determines the size of holonomy
corrections.\footnote{Specifically, it is defined in \cite{EffRecollapse} as
  $\delta = \frac{\sqrt{\Delta}}{2}$ with $\Delta \equiv 2\sqrt{3} \pi \gamma
  \ell_{\rm P}^2$ referring, as in \cite{APSII}, to the smallest non-zero area
  eigenvalue in loop quantum gravity. Owing to the lack of contact between
  loop quantum cosmology and the area spectrum in loop quantum gravity, we
  refrain from fixing $\delta$ and rather keep it as a free quantum
  parameter. Constructions of inhomogeneous models \cite{InhomLattice} suggest
  that $\delta$ is not related to the general area spectrum but instead to
  detailed properties of the quantum state underlying quantum space-time.}
Looking at large-scale evolution of large $|p|$, we are free to fix the sign
of $p$ or $\nu$, from now on assumed positive.  (These variables are a
specific case of the general parameterization of \cite{InhomLattice},
providing different cases of lattice refinement. While the harmonic model has
a dynamics independent of the specific choice of basic variables where $\nu$
could be any power of $p$, quantum corrections in models that differ for
instance by factor ordering depend on such parameters. Here, however, we
confine ourselves to the choice made in \cite{EffRecollapse} because our aim
is to explore the scenario suggested there.)

In the new variables, the classical Hamiltonian constraint is
\begin{equation}
\label{constraint}
- \frac{ 9 \, \nu  b^2} {16 \pi G \gamma^2 \delta } + 
\frac{ p^2_\phi } {3 \delta \, \nu} = 0.
\end{equation}
We use the free scalar $ \phi $ as an internal time variable; the monotonicity
of $\phi ( \tau )$ is guaranteed for the free case, and so evolution for $b
(\phi)$ and $ \nu (\phi)$ is governed by the Hamiltonian $p_\phi (b, \nu)$
which is found as a phase-space function by solving for $p_\phi$ in
Eq.~(\ref{constraint}). Doing this, we have $p_\phi \propto |\nu b|$. (In
  Sec.~\ref{factor ordering} we discuss ambiguities related to the choice of
  internal time.)

The momentum $p_{\phi}(\nu,b)$ of internal time, as a function of the
remaining degrees of freedom, acts as the Hamiltonian generating
evolution. Since it is quadratic, the system is equivalent to a harmonic one,
which, like the harmonic oscillator, can be quantized to a model without
quantum back-reaction.\footnote{The absolute value does not spoil this feature
  \cite{BounceCohStates} when used in effective equations: $\nu b$ as a
  Hamiltonian is $\phi$-independent and thus preserved. If one makes sure that
  an initial state is supported only on the positive part of the spectrum of
  $\nu b$, it will remain so supported. On the evolved state the action of the
  quantized $\nu b$ then agrees with that of the quantized $|\nu b|$.}
However, in a model of quantum gravity there may be quantum-geometry
corrections as well, changing the Hamiltonian not just by quantum
back-reaction terms but also by more-severe modifications.  In loop quantum
cosmology, as one popular example, holonomy and inverse-triad
corrections occur.

Holonomy corrections replace the $b$ in $p_{\phi}(\nu,b)$ by the non-linear
$\sin(b)$, making the Hamiltonian non-quadratic in canonical variables.
Nevertheless, as shown in \cite{BouncePert}, surprisingly, they do not remove
harmonicity even upon quantization provided a suitable factor ordering is
chosen: If we define $J:=\nu\exp(-ib)$, the new set $(\nu,J)$ of
(non-canonical) basic variables satisfies a linear algebra under Poisson
brackets, in which the {\em linear} Hamiltonian $p_{\phi}=|{\rm Im}(J)|$ is
just the right holonomy-modified version of the internal-time Hamiltonian
$p_{\phi}$. These properties are preserved when the model is quantized with
Hamiltonian $\hat{p}_{\phi}= |\frac{1}{2}i(\hat{J}-\hat{J}^{\dagger})|$, in this
specific ordering when seen as a function of $b$ and $\nu$. With a linear
Hamiltonian in variables obeying a linear algebra, no back-reaction
occurs. (Again, the absolute value turns out not to spoil harmonicity
features.)

Unlike the harmonic oscillator in quantum mechanics, the quadratic Hamiltonian
here is subject to factor ordering choices, and only one of them, as remarked
in \cite{BouncePert}, results in an exactly solvable quantum system as just
described. At the quantum level, $\hat{p}_{\phi}=|{\rm Im}\hat{J}|=
|\frac{1}{2}i(\hat{J}-\hat{J}^{\dagger})|$ writes the modified $|\nu\sin(b)|$
in the specific form $|\frac{1}{2}i(\nu\exp(-ib)-\exp(ib)\nu)|$. (In the
Wheeler--DeWitt limit $\delta\to0$ in which holonomy corrections disappear,
the ordering reduces to the standard symmetric one,
$\frac{1}{2}|\hat{\nu}\hat{b}+\hat{b}\hat{\nu}|$.) Any other choice gives rise
to quantum back-reaction, as spelled out in more detail below.  Moreover,
there is a second type of quantum-geometry modifications --- inverse-triad
corrections \cite{QSDV,InvScale,LoopMuk} --- which implies quantum
back-reaction even if the original ordering is chosen in which holonomy
corrections alone would not spoil harmonicity. (In this case, the $\nu$-
rather than $b$-dependence of the classical Friedmann equation is modified.)

\subsection{Effective equations and factor ordering}

To derive effective equations from our classical Hamiltonian above, we follow
the general procedure of \cite{EffAc,Karpacz}. For semiclassical (or even more
general) states, the method of effective equations allows us to avoid the
technical difficulties of working directly with wave functions and
operators. Also the hard problem of deriving physical Hilbert spaces and
explicit representations for constrained systems is avoided, and yet physical
normalization is implemented by reality conditions \cite{EffCons}. Large
classes of states can be dealt with by suitable parameters, such as
expectation values and fluctuations, of direct statistical significance for
physical properties. Effective equations therefore allow for much larger
generality, and thereby more reliable conclusions, than traditional methods
used in quantum cosmology. Our analysis of factor-ordering ambiguities
provides a further example.

Instead of working with wave functions, we have a framework in which states
are represented by the expectation values and moments they imply for basic
operators. As quantum variables in addition to expectation values, we use the
moments which describe a general quantum state:
\begin{equation}
G^{\underbrace{\scriptstyle \nu \, \cdots \, \nu}_m\underbrace{\scriptstyle b
    \, \cdots \, b}_n}
=\langle(\hat{\nu}-\langle\hat{\nu}\rangle)^m 
(\hat{b}-\langle\hat{b}\rangle)^n\rangle_{\rm Weyl}
\label{eq:moments}
\end{equation}
where $m,n$ are positive integers such that $m+n \geq 2$ and the subscript
Weyl denotes totally symmetric ordering of the operators. The moments and
expectation values define a phase space with Poisson bracket
\begin{equation}\label{Poisson}
\{\langle\hat{A}\rangle,\langle\hat{B}\rangle\}=
\frac{\langle[\hat{A},\hat{B}]\rangle}{i\hbar}
\end{equation}
in terms of the commutator, extended
by the Leibniz rule to arbitrary polynomials of the expectation values as they
occur in moments. Semiclassical regimes can be defined very generally to any
integer order $N$, by keeping moments only up to order $2N$. Indeed, in a
Gaussian state, the prime example of a semiclassical one, the moments behave
as
\[
 G^{\underbrace{\scriptstyle \nu \, \cdots \,
    \nu}_m\underbrace{\scriptstyle b \, \cdots \, b}_n}=:G^{m,n}\sim
\hbar^{(m+n)/2}\,.
\] 
However, our semiclassical regimes defined by the order of moments are much
more general than the 1- or at most 2-parameter families of Gaussians, and
thus avoid possible artifacts due to the selection of states.

The basic identity ${\rm d}\langle\hat{A}\rangle/{\rm d}t=
\langle[\hat{A},\hat{H}]\rangle/i\hbar$ of quantum mechanics, used for
instance to derive Ehrenfest's equations, can then be written as a
classical-type Hamiltonian flow generated by the quantum Hamiltonian
$H_Q(\langle\hat{\nu}\rangle, \langle\hat{b}\rangle,
G^{m,n})=\langle\hat{H}\rangle_{\langle\hat{\nu}\rangle,
  \langle\hat{b}\rangle, G^{m,n}}$ where the expectation value is computed in
a state specified by expectation values $\langle\hat{\nu}\rangle$ and
$\langle\hat{b}\rangle$ and all its moments $G^{m,n}$. Such expectation values
of interacting Hamiltonians may be difficult to compute, but in semiclassical
(or other) regimes in which only finitely many moments matter, they can be
derived perturbatively by expansions such as
\begin{eqnarray}
H_Q (\nu, b, G^{m,n}) & \equiv & \langle \hat{H} \rangle = \langle
H(\hat{\nu}, \hat{b}) \rangle = \langle H(\langle \hat{\nu} \rangle + (
\hat{\nu} - \langle \hat{\nu} \rangle), \langle \hat{b} \rangle + ( \hat{b} -
\langle \hat{b} \rangle )) \rangle\nonumber \\ 
\label{HQ} & = & H( \langle \hat{\nu} \rangle, \langle \hat{b} \rangle) +
\sum_{m,n: m+n \geq 2} \frac{1}{m!n!} \frac{ \partial^{m+n} H( \langle
  \hat{\nu} \rangle, \langle \hat{b} \rangle) }{\partial \langle \hat{\nu}
  \rangle^m \partial \langle \hat{b} \rangle^n } G^{m, n}  \,.
\end{eqnarray}
(We define $ \nu \equiv \langle \hat{\nu} \rangle$ and $b \equiv \langle
\hat{b} \rangle$ as a short cut used from now on.) In this particular
expression, we assume the Hamiltonian operator $\hat{H}$ to be Weyl ordered
just as the moments. 

If there are reasons for working with a different ordering, as explored below,
it differs from the Weyl-ordered one by re-ordering terms which can be
expressed by the moments as well. Accordingly, there will be additional
quantum corrections in (\ref{HQ}). For instance, if $H(\nu,b)=\nu^2 b^2$, the
Weyl-ordered quantization would be $\hat{H}=\frac{1}{6}(\hat{\nu}^2\hat{b}^2+
\hat{\nu}\hat{b}\hat{\nu}\hat{b}+ \hat{\nu}\hat{b}^2\hat{\nu}+
\hat{b}\hat{\nu}^2\hat{b}+ \hat{b}\hat{\nu}\hat{b}\hat{\nu}+
\hat{b}^2\hat{\nu}^2)$. Another symmetric ordering is $\hat{H}'=
\frac{1}{2}(\hat{\nu}^2\hat{b}^2+\hat{b}^2\hat{\nu}^2)$. Using
$\hat{\nu}\hat{b}\hat{\nu}\hat{b}+ \hat{b}\hat{\nu}\hat{b}\hat{\nu}=
\hat{\nu}^2\hat{b}^2+\hat{b}^2\hat{\nu}^2- [\hat{b},\hat{\nu}]^2$ and
$\hat{\nu}\hat{b}^2\hat{\nu}+\hat{b}\hat{\nu^2}\hat{b} =
\hat{\nu}^2\hat{b}^2+\hat{b}^2\hat{\nu}^2 -2[\hat{b},\hat{\nu}]^2$ for
canonical commutators of $\hat{\nu}$ and $\hat{b}$, we have
$\hat{H}'=\hat{H}+\frac{1}{2}[\hat{b},\hat{\nu}]^2$. The effective
Hamiltonians
\[
 \langle\hat{H}\rangle= \nu^2b^2+b^2G^{\nu\nu}+4\nu bG^{\nu b}+ \nu^2G^{bb}+
 2bG^{\nu\nu b}+2\nu G^{\nu bb}+G^{\nu\nu bb}
\]
and $\langle\hat{H}'\rangle=\langle\hat{H}\rangle-\frac{1}{2}\hbar^2$ differ
just by constants, but for higher polynomials moment-dependent factor-ordering
terms can result.  We will present a more detailed example below.

In general, any ordering of a term $\widehat{\nu^m b^n}$ in a
polynomial (or series-expanded) $\hat{H}$ can be written as
\begin{equation}
 \widehat{\nu^m b^n}= \widehat{(\nu^m b^n)}_{\rm
   Weyl}+ \sum_{j=1}^{\min(m,n)} c_j \hbar^j \widehat{(\nu^{m-j} b^{n-j})}_{\rm
   Weyl}
\end{equation}
by a finite number of applications of $[\hat{b},\hat{\nu}]=i\hbar$. Effective
Hamiltonians of differently ordered Hamiltonian operators therefore differ
only in terms containing explicit factors of $\hbar$. Only even powers of
$\hbar$ occur since $[\hat{b},\hat{\nu}]=i\hbar$ must be applied an even
number of times to relate symmetric orderings, ensuring that the associated
effective Hamiltonians are real.

Quantum evolution --- determined from commutators with the Hamiltonian
operator --- is the Hamiltonian flow of $H_Q=\langle\hat{H}\rangle$,
generating equations of motion for expectation values and quantum moments:
${\rm d} \langle \hat{A} \rangle / {\rm d} t = \{ \langle\hat{A}\rangle , H_Q
\}$ using (\ref{Poisson}). Unless the classical Hamiltonian is quadratic in
canonical variables, the last term in (\ref{HQ}) contains products of
expectation values and moments. Their equations of motion are then coupled,
with moments back-reacting on the dynamics of expectation values as a source
of quantum corrections. Re-ordering terms added to (\ref{HQ}) for a non-Weyl
ordered Hamiltonian thereby affect the effective dynamics.

\subsection{Modifications in (loop) quantum cosmology}

In this paper, we will work to leading order $m+n=2$, and ignore higher
moments as they are subdominant for semiclassical states. (For analyses of
cosmological quantum back-reaction using this scheme with moments of different
orders, see \cite{BounceSqueezed,HighDens,HigherMoments}.)  Moreover, as we
will see below, they appear with factors which are suppressed by inverse
powers of $\nu$ and become even more suppressed at large volume.
Still, to leading order there is quantum back-reaction of fluctuations and
correlations in expectation values. To compare evolution equations with a
single Friedmann equation of the usual form, we should solve for all moments
included and insert the solutions in equations of motion for expectation
values. But as we are, for now, looking at particular terms in a general
formulation, we do not have an explicit system of equations to solve.

We are extending the typical analysis by including three forms of corrections
--- the holonomy replacement, factor ordering ambiguities, and inverse triad
corrections --- which we will now define and summarize. The holonomy
replacement is motivated from the full theory of loop quantum gravity
\cite{Rov,ThomasRev,ALRev}, as already discussed. Furthermore, holonomy
corrections are of interest here because we are looking to reproduce the $
\epsilon^2 $ term of Eq.~(\ref{ym friedmann}). In our framework,
$\frac{1}{2}\epsilon^2 = G^{bb}$ is the squared variance of $b$. Thus, if we
are looking for a $G^{bb}$ dependence in our corrections to the Friedmann
equation, we require $H$ to have a non-zero third derivative with respect to
$b$. (From Eq.~(\ref{HQ}) we see that to obtain a $G^{bb}$ term we need at
least a non-zero second derivative, and as we will see in more detail, we need
an additional derivative of $b$ to get $G^{bb}$ in the equations of motion
from the Poisson bracket.) The holonomy replacement easily satisfies this
requirement, but less obviously it may be realized also in
Wheeler--DeWitt quantum cosmology with a non-standard factor ordering.

Inverse-triad corrections modify the $\nu$-dependence of the Hamiltonian. The
Hamiltonian constraint (\ref{Friedmann}) to be quantized contains inverse
powers of $\nu$. Also here, as with holonomy corrections, the full theory
suggests characteristic modifications by elementary properties of the form of
quantum geometry realized. The variable $\nu$, or rather $p$ as the isotropic
component of the densitized triad used as a basic field in the full theory, is
quantized to flux operators with discrete spectra containing zero. Such
operators have no densely defined inverse, and therefore one must use more
indirect quantization techniques, provided in \cite{QSDI,QSDV} to represent
inverses of $\nu$ or $p$ in the Hamiltonian constraint. In isotropic models,
the resulting correction functions can be computed explicitly
\cite{InvScale,Ambig,ICGC}; they appear in quantized Hamiltonian constraints
and determine, for instance, the coefficients of dynamical difference
equations. For our present purposes, we take these effects into account by
writing the $\nu$-factor of $p_{\phi}$ as a general function $f(\nu)$ which,
in the large volume regime considered here, has the form $f(\nu)\sim \nu
(1+O(1/\nu))$. A commonly used parameterization, for instance in
phenomenological analysis, is $\nu + c ( \ell_\mathtt{P}^3 / \nu
)^{n}$ with $c$ real valued and $n$ a positive integer.

Summarizing holonomy and inverse-triad corrections, we find our Hamiltonian
with corrections to be
\begin{equation}
\label{hamiltonian}
H( b , \nu) \equiv p_\phi = \sqrt{ \frac { 27} {16 \pi G} } \frac{ f(\nu) \sin
  b} {\gamma} \,. 
\end{equation}
The next step is to use this example Hamiltonian to evaluate the dynamics of
our system and arrive at an effective Friedmann equation.

\subsubsection{Effective dynamics}

Following the effective-equation prescription, we find our
quantum Hamiltonian, up to second order in moments, to be
\begin{equation}
\label{hq}
H_Q = \frac{ 1} { \gamma} \sqrt{ \frac{ 27} {16 \pi G} }  \left[ f( \nu ) \sin
  b - \frac{1}{2} G^{bb} f( \nu ) \sin b + G^{b \nu} f'( \nu ) \cos b +
  \frac{1}{2} G^{\nu \nu} f''(\nu) \sin b \right] 
\end{equation}
where the prime represents a derivative with respect to $\nu$. As mentioned in
the general discussion of effective equations, the specific terms written here
assume that the Hamiltonian operator is Weyl ordered in $\hat{\nu}$ and
$\hat{b}$. At this stage, factor-ordering choices may introduce additional
quantum corrections which also depend on the moments, as discussed in more
detail below.

From (\ref{hq}), we can find the relational equations of motion for our
expectation values via Poisson brackets. For instance, we have the expectation
value of the volume in terms of expectation values and moments:
\begin{eqnarray}
\label{nudot}
\nonumber \frac{ {\rm d} \langle \hat{\nu} \rangle}
{{\rm d} \phi}   &=&  \{ \nu , H_Q \} \\
& = & - \sqrt{ 12 \pi G } \left[ f( \nu ) \cos b - \frac{ 1}{2} G^{bb}  f( \nu
  ) \cos b - G^{b \nu} f'(\nu) \sin b + \frac{1}{2} G^{\nu \nu} f''( \nu )
  \cos b \right]. 
\end{eqnarray}
Here our moments back-react on the classical trajectory. 
The moments are not constant and subject to their own equations of motion, but
we will not need the latter for the subsequent analysis.

\subsubsection{Friedmann equation}

Now let us find the quantum corrections to the Friedmann equation for our
example. From Eq.~(\ref{b and v}) and the definition of $|p| = a^2$, we have
\begin{eqnarray}
\label{dnu dtau}
\frac{ {\rm d} \nu} {{\rm d} \tau} & = &\frac{ 2 \, a^2} {\delta}
\frac{ {\rm d} a} { {\rm d} \tau} = \frac{ {\rm d} \nu} {{\rm d}
  \phi} \frac{ {\rm d} \phi} {{\rm d} \tau}  \\ 
\frac{ {\rm d} \phi} {{\rm d} \tau} & = & \frac{ 2 \, p_\phi } { 3
  \delta \, \nu } 
\end{eqnarray}
where the proper-time derivative ${\rm d}\phi / {\rm d}\tau$ is generated by
the original Hamiltonian constraint. (At low curvature, we need not worry
about signature change \cite{Action,SigChange} which could eliminate time.)
Rearranging and substituting, we find
\begin{eqnarray}
\label{friedmann1}
\left( \frac{ 1} { a} \frac{ {\rm d} a} {{\rm d} \tau} \right)^2 & = & \frac
{4} {81 
  \delta^2}  \left( \frac{ {\rm d} \nu/{\rm d}\phi }{ \nu} \right)^2 \left(
\frac{ p_\phi}{     \nu} \right)^2 \\ 
\label{friedmann} & = & \frac{4}{81 \delta^2} \left(12 \pi G \, \frac{ f(\nu)^2}
{\nu^2} \cos^2 b \, \left( 1 - G^{bb} \right)\right)  \left(\frac{
  27} {16 \pi G\gamma^2} \frac{ f(\nu)^2} {\nu^2} \sin^2 b  (1-G^{bb})\right)
\end{eqnarray}
where we used Eqs.~(\ref{nudot}) and (\ref{hq}), now with $p_{\phi}=H_Q$, to
arrive at the second line and dropped squared moments as well as the $f'$ and
$f''$ terms for simplicity. Note that these derivative terms are not neglected
in the actual analysis to follow. In terms of energy density we have
\begin{equation}
\label{rho}
\rho = \frac{ H_{\rm matter} }{a^3} = \frac{2} {9 \delta^2} \frac{ p^2_\phi} {
  \nu^2} = \frac{3} {8 \pi G\gamma^2\delta^2} \frac{ f(\nu)^2} {\nu^2}
\sin^2 b (1-G^{bb})
\end{equation}
where we again used Eq.~(\ref{hq}) for $H_Q=p_{\phi}$. Rearranging to solve
for $\sin^2 b$, we can substitute into Eq.~(\ref{friedmann}) to find
\begin{eqnarray}
\label{friedmann real}
\left( \frac{ 1} { a} \frac{ {\rm d} a} {{\rm d} \tau} \right)^2 &\sim& 
\frac{ 8 \pi G} {3}
\rho \left( \frac{ f(\nu)^2} {\nu^2} - \frac{8 \pi G\gamma^2}{3} \delta^2
   \rho (1+G^{bb}) \right) \left( 1 - G^{bb} \right)\nonumber\\
& \sim& \frac{8\pi G}{3}\rho
 \left(\frac{f(\nu)^2}{\nu^2}- \frac{8\pi G\gamma^2}{3}\delta^2
   \left(\rho+\frac{3f(\nu)^2}{8\pi G\gamma^2\delta^2\nu^2}
     G^{bb}\right)\right) \,. 
\end{eqnarray}
The fluctuation
added to $\rho$ in the combination
\[
 \rho_{\rm Q}=\rho+\frac{3f(\nu)^2}{8\pi G\gamma^2\delta^2\nu^2} G^{bb}
\]
is, within the approximations used here, analogous to the correction found in
\cite{QuantumBounce,BounceSqueezed} (see also \cite{QCEffCons}).

\section{General factor ordering and collapse terms}

In addition to the holonomy replacement and inverse-triad corrections, we
include the possibility of a factor-ordering ambiguity as the main ingredient
in this paper. None of these corrections are expected to be very large in
low-curvature semiclassical regimes, but they could play a role in long-term
evolution. To show that quantum cosmology has the correct large-scale
behavior, all these terms must be studied carefully, but no such analysis has
been completed yet. Here, we present systematic expansions by which the issue
of long-term behavior can be addressed, focusing specifically on the
question of whether semiclassical effects can be so strong that they trigger a
recollapse of a classically ever-expanding universe.

\subsection{Factor ordering}
\label{factor ordering}

The Hamiltonian constraint of (loop) quantum cosmology (or loop quantum gravity
in general) is far from being unique, owing for instance to choices in the
representation of holonomies and factor orderings. Although uniqueness claims
seem to exist in the literature, they are based on certain mathematical
features posed ad-hoc, rather than physical considerations; they cannot be
used when addressing the semiclassical limit, long-term evolution, or concrete
physical effects.

In addition to ambiguities in the formulation of the Hamiltonian-constraint
operator, further choices are usually made when one addresses the problem of
time. One chooses or even introduces simple matter degrees of freedom, here
and elsewhere assumed as a free and massless scalar field, whose momentum is
exactly constant and, in particular, never vanishing. The scalar degree of
freedom therefore has no turning points when solved for as a function of
proper time, and can be used as a mathematical evolution parameter itself.

Simple and manageable equations at the classical or quantum levels usually
result. However, the question of whether different choices of internal times
provide the same physics remains complicated to address; see e.g.\
\cite{ReducedKasner,Electric}. (An exception is given by semiclassical
regimes, where effective methods allow one to change time by gauge
transformations of moments \cite{EffTime,EffTimeLong,EffTimeCosmo}.) If
independence of one's choice of internal time can be achieved at all, as some
kind of space-time anomaly freedom, it likely requires detailed prescriptions
of factor orderings and other quantization ambiguities. The Hamiltonian one
may choose in a deparameterizable version in rather simple terms may not be
what is required for the correct physics. Therefore, a wider view on factor
ordering and other choices is preferable for reliable statements, implemented
here by general parameterizations.

Instead of directly quantizing $\nu\sin(b)$ with rather controlled factor
ordering options, or in the harmonic formulation the even more unique-looking
${\rm Im}J$, we should start with the Hamiltonian constraint
(\ref{modconstraint}) and its terms $p_{\phi}^2/\nu$ and $\nu\sin^2(b)$. The
types of factor-ordering ambiguities differ in those two cases, showing the
restricted nature of deparameterized models regarding factor ordering. It is
not easy to parameterize all possible ordering effects, and therefore we will
present the main implications by key examples.

In the former case, looking at possible orderings of the $\phi$-Hamiltonian
$\nu\sin(b)$, we could take a symmetric ordering such as
$\frac{1}{2}(\hat{\nu}\widehat{\sin(b)}+ \widehat{\sin(b)}\hat{\nu})$ --- the
Weyl ordering of operators $\hat{\nu}$ and $\widehat{\sin(b)}$ --- but also
$\sqrt{\hat{\nu}}\widehat{\sin(b)}\sqrt{\hat{\nu}}$ or a more complicated
re-ordering. The first two differ by $\frac{1}{2}
(\sqrt{\hat{\nu}}[\widehat{\sin(b)},\sqrt{\hat{\nu}}]+
[\sqrt{\hat{\nu}},\widehat{\sin(b)}]\sqrt{\hat{\nu}})=
\frac{1}{2}[\sqrt{\hat{\nu}},[\widehat{\sin(b)},\sqrt{\hat{\nu}}]]$. Such
terms, in a low-curvature expansion, just add positive powers of $b$ with
ordering-dependent coefficients. The classical correspondence of the ordering
term just derived, for instance, is
\begin{equation} \label{reord}
-\frac{1}{2}\hbar^2\{\sqrt{\nu},\{\sin(b),\sqrt{\nu}\}\}=
-\frac{1}{9}\pi^2\gamma^2\ell_{\rm P}^4\nu^{-1} \sin(b)\,.
\end{equation}
As in this example, iterated commutators, or Poisson brackets in the classical
correspondence, amount to terms of the order $\ell_{\rm P}^4/\nu$ to some
positive power, by which various orderings differ. (A single commutator
contributes a factor of $i\ell_{\rm P}^2/\sqrt{\nu}$. However, there must
always be an even number of commutators if the difference of two symmetric
orderings is considered to avoid factors of $i$.) Such terms can be
interpreted as changing the form of inverse-triad corrections in $f(\nu)$ or
the choice of holonomy corrections initially implemented by the sine function,
and therefore do not differ much from the usual ambiguities of holonomy
corrections.

Factor-ordering ambiguities can, however, be more radical if we start with the
original Hamiltonian constraint rather than the deparameterized
$\phi$-Hamiltonian. One could first take the square root of the quantized
$p_{\phi}^2/\nu\propto \nu\sin^2(b)$, for instance of
$\hat{\nu}\widehat{\sin(b)}^2+\widehat{\sin(b)}^2\hat{\nu}$, and then multiply
with $\sqrt{\hat{\nu}}$ to obtain a version of
$\hat{p}_{\phi}$. Factor-ordering choices now include expressions such as
$\hat{\nu}^{1/4}\sqrt{\hat{\nu}\widehat{\sin(b)}^2+
  \widehat{\sin(b)}^2\hat{\nu}}\: \hat{\nu}^{1/4}$ which, compared with the
previous ones, differ by terms that require some quantization of the inverse
$(\nu\sin(b)^2+\sin(b)^2\nu)^{-1/2}$. It is more diffcult to compute
commutators of square-root operators to see what factor-ordering terms now
arise.

However, we can estimate contributions in an
$\hbar$-expansion, suitable for effective equations, by noting that arguments
of the square root such as $\hat{\nu}\widehat{\sin(b)}^2+
\widehat{\sin(b)}^2\hat{\nu}$ differ from squares of our previous Hamiltonians
by double commutators as in (\ref{reord}), but involving a power of $\nu$ and
$\sin^2(b)$ (not $\sin b$ as before). For instance, comparing the two
orderings $\hat{\nu}\widehat{\sin(b)}^2+\widehat{\sin(b)}^2\hat{\nu}$ and
$2\sqrt{\hat{\nu}}\widehat{\sin (b)}^2\sqrt{\hat{\nu}}$, we have 
\begin{eqnarray}
&&\hat{\nu}\widehat{\sin(b)}^2+\widehat{\sin(b)}^2\hat{\nu}-
2\sqrt{\hat{\nu}}\widehat{\sin (b)}^2\sqrt{\hat{\nu}}=
\left[\sqrt{\hat{\nu}},\left[\sqrt{\hat{\nu}},
\widehat{\sin(b)}^2\right]\right]\nonumber\\ 
&\sim&
-\frac{\hbar^2}{2\nu} (\sin(b)^2-\cos(b)^2)= -\frac{\hbar^2}{2\nu}\cos(2b)\,.
\label{double}
\end{eqnarray}
In the semiclassical correspondence, such a re-ordering term added to
$\nu \sin^2(b)$ under the square root implies a leading $\hbar$-correction of
\begin{equation} \label{root}
\sqrt{\nu\sin^2(b)-\frac{1}{2}\hbar^2\nu^{-1}\cos(2b)}- \sqrt{\nu} \sin(b)=
-\frac{\hbar^2}{4\nu^{3/2}}\frac{\cos(2b)}{\sin(b)}+
O(\hbar^4/(\nu^{7/2}\sin(b)^3))\,,
\end{equation}
which is not analytic at $b=0$. (Semiclassical physics follows from an
expansion in $\hbar$, assumed to be small. But when $b$ is small too, the
square-root is evaluated near its non-analytic point.) Such inverse-power
terms will give us more options for moments in an effective Hamiltonian
possibly resembling the one used in \cite{EffRecollapse}.

If we include inverse-triad corrections by a generic $f(\nu)$ instead of
$\nu$, we will have double commutators of the form $[\sqrt{f(\hat{\nu})},
[\sqrt{f(\hat{\nu})},\sin^2(b)]]$. The $1/\nu$ in (\ref{double}) will then be
replaced by $f'(\nu)^2/f(\nu)$. Taking the square root and expanding as in
(\ref{root}) will produce a series
\begin{eqnarray} \label{rootexpand}
 \sqrt{f(\nu)\sin^2(b)-\frac{1}{2}\hbar^2\frac{f'(\nu)^2}{f(\nu)}
   \cos(2b)}&\sim& \sqrt{f(\nu)}\sin(b)\\
&&- \frac{1}{4}\hbar^2
 \frac{f'(\nu)^2}{f(\nu)^{3/2}} \frac{\cos(2b)}{\sin(b)}+
 O(\hbar^4f'(\nu)^4/(f(\nu)^{7/2}\sin(b)^3))\,. \nonumber
\end{eqnarray}
This series would be used as the gravitational contribution to the effective
Hamiltonian constraint, which is to equal the effective matter contribution
$p_{\phi}^2/\nu$. Also here, in the matter contribution, inverse-triad
corrections $g(\nu)p_{\phi}^2$ may result, in general by a function
$g(\nu)\sim 1/\nu(1+O(1/\nu))$ that differs from $1/f(\nu)$ in higher-order
terms. Assuming the ordering just discussed, the effective $p_{\phi}$
generating $\phi$-evolution is then
\begin{eqnarray} \label{Hexp}
 p_{\phi}&=&
 \frac{\sqrt{\langle\widehat{f(\nu)\sin^2(b)}\rangle}}{\sqrt{g(\nu)}}\\  
&\sim & \frac{1}{\sqrt{f(\nu)g(\nu)}} \left(f(\nu)\sin(b)- \frac{1}{4}\hbar^2
 \frac{f'(\nu)^2}{f(\nu)} \frac{\cos(2b)}{\sin(b)}+
 O(\hbar^4f'(\nu)^4/(f(\nu)^{3}\sin(b)^3))\right)\,. \nonumber
\end{eqnarray}

The types of ordering just presented are more involved than simple ones when
one directly quantizes $\nu\sin(b)$. However, they are no less natural; in
fact, they could be argued to be more natural from the point of view that the
terms in the original Hamiltonian constraint, $1/\nu$ and $\nu\sin^2(b)$,
should first be quantized, with a square root for $p_{\phi}$ taken afterwards
at the quantum level. For full generality and to guarantee independence of
one's results from special features available only in deparameterized systems
but not otherwise one must take the properties of the orderings shown here
into account. This realization makes the low-curvature regime of quantum
cosmology much more complicated than it may naively seem.

In a low-curvature expansion of expressions such as (\ref{root}), even
inverse powers of $b$ can appear due to factor ordering. This feature may
sound surprising because such terms seem unduly large in semiclassical
regimes. However, they are always accompanied with factors of $\hbar$ and, in
some cases, moments such as the curvature fluctuation which decrease as $b$
gets small. The fate of long-term evolution now crucially depends on
properties of state parameters and the detailed form of quantum corrections
and effective equations, which requires a dedicated analysis on which we now
embark.

\subsection{Low-curvature expansion}

So far, we have kept all terms expected from loop quantum cosmology, but they
will not be fully required in the semiclassical regimes we are interested
in. In the small-curvature regime, we can expand the Hamiltonian in powers of
$b$:
\begin{equation}
\label{general h}
H = \sum_{n\in{\mathbb Z} \:{\rm odd}}H_n(\nu)b^n= H_1(\nu) b + H_3(\nu) b^3
+\cdots + H_{-1}(\nu) b^{-1}+\cdots 
\end{equation}
where the $H_{n}$ are functions of $\nu$ only. Such an expansion allows us to
consider generic orderings.  We include inverse powers of $b$ as they may
arise from factor orderings; these terms therefore all include factors of
$\hbar$. (After expanding all trigonometric functions in (\ref{Hexp}), also
the positive-power terms in the $b$-expansion will be corrected by reordering
terms with explicit factors of $\hbar$.) They will appear explicitly in the
effective Hamiltonian, added to the original (\ref{HQ}) which is based on an
operator in which all contributions are Weyl ordered. The effective
Hamiltonian will contain moments for all terms in the curvature expansion,
including the inverse-power ones.  At small curvature, these latter terms will
become progressively more important. We will illustrate our main analysis
based on the truncation shown in (\ref{general h}), which will not allow us to
address the recollapse (where $b=0$). However, our approach will be sufficient
to investigate the possible terms that can appear in a general effective
Friedmann equation for a model without ad-hoc assumptions.

In (\ref{general h}), the linear term $H_1b$ is what is expected
classically. The cubic term is the first higher-order quantum correction due
to the holonomy modification; even higher orders are not relevant at low
curvature.  For very small and very large curvature, respectively, we should
expand to higher orders of positive and negative powers in $b$, but as we will
discuss in the analysis, this does not lead to categorically different terms.

{}From (\ref{general h}), we can then solve for the equation of motion for $\nu$
\begin{equation}
\label{general nudot}
\dot{\nu} = \{ \nu , H_Q \} = \{ \nu , H \} + \frac{1}{2} G^{bb} \{ \nu
, \partial^2_b H \} + G^{b \nu} \{ \nu , \partial_b \partial_\nu H \} +
\frac{1}{2} G^{\nu \nu} \{ \nu , \partial^2_{\nu} H \} 
\end{equation}
to second order in moments, which we expand term by term to find
\begin{eqnarray}
\label{ nu pb h}
\{ \nu, H \} & = & - \alpha H_1 - 3 \alpha H_3 b^2 + \alpha H_{-1} b^{-2} \equiv A \\ 
\frac{1}{2} G^{bb} \{ \nu , \partial^2_b H \} & = & \frac{1}{2} G^{bb}
\,  \partial^2_b A = \frac{1}{2} G^{bb} \left( -6 \alpha H_3 + 6 \alpha H_{-1}
  b^{-4} \right) \\ 
G^{b \nu} \{ \nu , \partial_b \partial_\nu H \} & = & G^{b \nu}
\, \partial_b \partial_\nu A = G^{b \nu} \left( - 6 \alpha H'_3 b - 2
  \alpha H'_{-1} b^{-3} \right) \\ 
\frac{1}{2} G^{\nu \nu} \{ \nu , \partial^2_{\nu} H \} & = & \frac{1}{2}
G^{\nu \nu} \partial^2_{\nu} A = \frac{1}{2} G^{\nu \nu} \left( - \alpha
  H''_{1} - 3 \alpha H''_{3} b^2 + \alpha H''_{-1} b^{-2}  \right) 
\end{eqnarray}
where $\alpha = 8 \pi G \gamma / 3$, $H'_{n} = \partial_\nu H_{n}$, and we
define $A$ in the first equation.

Now we have ${\rm d}\nu/{\rm d}\phi$ expanded in powers of $b$. Recall
however, that for the Friedmann equation, we are interested in
\begin{equation}
\label{friedmann2}
\left( \frac{ 1} { a} \frac{ {\rm d} a} {{\rm d} \tau} \right)^2 =  \frac { 4}
{81 
  \delta^2}  \left( \frac{ {\rm d}\nu/{\rm d}\phi }{ \nu} \right)^2 \left(
  \frac{ p_\phi}{ 
    \nu} \right)^2 = \frac { 4} {81 \delta^2} \left( \frac{ \{ \nu , H_Q
    \} } { \nu} \right)^2 \frac{ H_Q^2 } { \nu^2}  
\end{equation}
which requires us to look at the expansion of $({\rm d}\nu/{\rm d}\phi)^2
H_Q^2/ \nu^4$. For terms of interest, those that depend on the second-order
moments but are independent of $\rho$, we can limit our search to terms with
$b^0$ and $\nu^0$ which would resemble the term responsible for a potential
recollapse. (This term, specifically, contains $G^{bb}$ but no factor of $\nu$
or $b$. There may be a small dependence on $\nu$, but it cannot be strong at
large volume because: If there is a power-law dependence with positive
exponent, the term would violate the semiclassical limit; if the exponent is
negative, the term will be suppressed at large volume.)  It turns out that the
dependence on expectation values is already a restrictive condition,
irrespective of the dependence on moments. From now on, we can ignore the
factor of $1/\sqrt{g(\nu)f(\nu)}$ in (\ref{Hexp}) because (in quadruplicate)
it simply multiplies the right-hand side of (\ref{friedmann2}) without
changing the leading-order dependence on $\nu$.

The simplest procedure is to look at all $b$-independent terms in $({\rm
  d}\nu/{\rm d}\phi)^2H_Q^2$ and then restrict to all remaining ones
proportional to $\nu^4$ to cancel the explicit $\nu$-dependence in
(\ref{friedmann2}). Since both ${\rm d}\nu/{\rm d}\phi= \{ \nu, H_Q \}$ and
$H_Q$ itself are expanded as in (\ref{general h}), all expansion terms in the
product are quartic expressions in the $H_n$. Terms independent of $b$ but
containing $G^{bb}$ include, for instance,
\begin{equation}
6 \alpha^2 G^{bb} \, G^{\nu \nu} H''_1 H_3 H_{1} H_{-1} \quad\mbox{or}\quad
18 \alpha^2 G^{bb} G^{bb} H_3 H_3  H_1 H_{-1} \,.
\end{equation}
In the first example, the contribution $G^{\nu \nu} H''_1$ comes from the
last term in (\ref{general nudot}), $G^{bb}H_3$ from the second term in the
same equation (both factors appearing in one term of $\dot{\nu}^2$), while
$H_1H_{-1}$ is one term in $H_Q^2$ that does not depend on $b$.  

Now we are interested in what is necessary such that the combination of $H(
\nu )$ gives us terms with $ \nu^0$, thus satisfying the $\rho$ independence,
which could dominate the Friedmann equation in a regime with low energy
density. With that in mind, and looking at (\ref{Hexp}), let us categorize the
$\nu$-dependence in the $H_n$ terms as
\begin{equation}
H_1 \sim f ( \nu ) \sim H_3 \quad,\quad
H_{-1} \sim \frac{ f' (\nu)^2 }{ f( \nu)}\quad,\quad
H_{-3} \sim \frac{ f' (\nu)^4 }{f (\nu)^3} \,.  
\end{equation}
where these associations come from the power expansion of our original $H$:
$H_1 \sim H_3$ are the coefficients of $ \sin b$, $H_{-1}$ and $H_{-3}$ arise
from the factor ordering implementation, each additional inverse power of $b$
requiring another commutator of $\widehat{\sin(b)}^2$ with $\widehat{f(\nu)}$
as in (\ref{rootexpand}). Recall that $f (\nu)$ is due to the inverse triad
corrections, $f(\nu)/\nu\sim \langle\hat{\nu}\rangle
\langle\widehat{\nu^{-1}}\rangle \approx 1 + O(1 / \nu )$. Although we have an
expectation for the form of $f(\nu)$ we do not presume it here; rather, we
will solve for the necessary values that provide $\rho$ independence and
compare for consistency.

With these classifications, we can then introduce a simple expression for
the combinations
\begin{equation}
\label{bbh}
\mathbb{H}_{S} \equiv H_A H_B
H_C H_D \sim f(\nu)^S f'(\nu)^{4-S} \,,
\end{equation}
depending only on the integer $S=\min(A,1) + \min(B,1) + \min(C,1) +
\min(D,1)$ for all possible values of $A$, $B$, $C$ and $D$.  (Note that from
above $H_1 \sim H_3$, a relation which extends to all higher-curvature terms
$H_{2n+1}$ with $n>0$ obtained from expanding $\sin(b)$. For this reason, we
use the minima of the order and one to compute the power of $f'(\nu)$. Each
$H_n$ then contributes $\min(n,1)$ factors of $f(\nu)$ and $1-\min(n,1)$
factors of $f'(\nu)$.) The maximum of $S$ according to its definition is
$S\leq 4$ which would be attained if and only if $A$, $B$, $C$ and $D$ are all
positive. However, this case cannot eliminate all factors of $b$ that come
along with the $H_n$. There must be at least one negative coefficient in
allowed terms, lowering the maximum to $S\leq 2$. This upper bound will play a
crucial role in our subsequent arguments.

We are then interested in $ \mathbb{H}_{S} $ such that we have a $ \nu^4$
dependence which cancels in Eq.~(\ref{friedmann2}). Thus we can solve for
$f(\nu)$ with respect to $\nu$ to see what is required for a $\nu$-independent
term in the effective Friedmann equation.  We need $f(\nu)^S f'(\nu)^{4-S}
\propto \nu^4$ or, ignoring for simplicity higher-order terms in a general
power law $f(\nu)\sim \nu^x(1+O(1/\nu))$, we must have $x=(8-S)/4$ or $f(\nu)
\sim \nu^{(8-S)/4}$.  Since $S$ is an integer at most 2 inclusive due to the
possible combinations of our terms, the smallest possible exponent is
$f(\nu)\sim \nu^{3/2}$ increasing by $\nu^{\, 1/4}$ as $S$ decreases in
integer steps, reaching a maximum of $f(\nu) \sim \nu^4$ for $S=-8$ (for the
truncation used here). Allowing for higher orders in the $b^{-1}$-expansion
(\ref{general h}) will only increase the power of $\nu$ in $f(\nu)$. None of
these behaviors is compatible with the classical limit $f(\nu)=\nu$.  Our
expectation of $f(\nu)$ due to inverse triad corrections is only compatible
with $f(\nu) \sim \nu$ in the case that there are no corrections or when $\nu$
is extremely large. However, the $S=4$ case is never realized as it would
require a four-term combination of only $H_1$ and $H_3$ that still gives $b^0$
in the effective Friedmann equation. We do have terms with only these two
factors in $H_Q^2$ but they correspond to positive, even powers of $b$ (in
fact, $H_A H_B$ corresponds to terms with $b^{A+B}$).  Similarly, terms with
only these two factors from $\dot{\nu}^2$ also correspond to positive or zero
powers of $b$ and thus the only way we could achieve a $\nu$-independent
correction would require a $b^2$ or higher dependence. For all other values of
$S$, we would require some $f(\nu) \sim \nu^x$ with $x \geq 3/2$ which, as
mentioned above, is not compatible with our requirements on $f(\nu)$.

Now let us return to the inclusion of higher derivative terms of $f(\nu)$ and
their effects on our analysis.  Returning to our previous notation (and using
$H_1 \sim H_3$ again) we have:
\begin{eqnarray}
\partial_\nu H_n & = & n H_{n-1} + (1-n) H_n \frac{ f''(\nu)}{f'(\nu)} \\
\nonumber \partial^{\, 2}_\nu H_n & = & (1-n) \left( -n H_n \left( \frac{ f''(\nu)}{f'(\nu)} \right)^2 + H_n \frac{f'''(\nu)}{f'(\nu)} \right) + \\ 
& & n \left( (2 - n) H_{n-1} \frac{ f''(\nu)}{f'(\nu)} + (1 - n) H_{n-1} \frac{ f''(\nu)}{f'(\nu)} \right) + n (n-1) H_{n-2} 
\end{eqnarray}
With the benefit of foresight, it is now prudent to extend Eq.~(\ref{bbh}) to
allow for more than four terms by defining
\begin{equation}
\mathbb{H}_{S=A+B+C+\cdots \, , \, N} = \overbrace{H_A H_B H_C \cdots }^{N \:{\rm
    terms}}  \sim f(\nu)^S f'(\nu)^{N -S}
\end{equation}
where $S$ is the sum of the labels as defined before and $N$ determines how
many $H_n$ terms are multiplied together. While we will continue to focus on
the $N=4$ case corresponding to our possible contributions to the effective
Friedmann equation, this parameter is important for categorizing
derivatives. In fact, with this definition, we can then write the solution to
terms which have derivatives as
\begin{equation}
\label{hn'}
\mathbb{H}^{\, \, n'}_{S  , \, N} = n \, \mathbb{H}_{S-1  , \, N} + (1 - n) \mathbb{H}_{ S , \, N-1} f''(\nu) 
\end{equation}
where the $n'$ superscript represents that there is a $H'_n$ term in the
combination and we see that $N$ is altered in terms which have higher
derivatives. We could also write out a general formula for combinations with
more than one derivative term; however, terms which include two derivative
factors are included (up to numerical coefficients), and
combinations that contain more derivative terms lead to inconsistent
conditions on $f(\nu)$, as we will see in Eq.~(\ref{generalized ansatz}).

We can also express terms with second derivatives:
\begin{eqnarray}
\nonumber \mathbb{H}^{ \, \, n''}_{S, \, N} & = & n (n-1) \mathbb{H}_{S-2, \, N} + n (3 - 2n) \mathbb{H}_{S-1 , \, N-1} f''(\nu) + \\
& & n (n-1) \mathbb{H}_{S , \, N-2} f''(\nu)^2 + (1-n) \mathbb{H}_{S , \, N-1} f'''(\nu)
\end{eqnarray}
for $n\leq 1$.  Now that we have these expressions, we once again must solve
for $f(\nu)$ such that we have a $\nu$-independent term in the effective
Friedmann equation. These solutions are obtained from differential equations
with non-linear products of varying orders of derivatives. We can generalize
these differential equations from our terms as:
\begin{equation} 
 \mathbb{H}_{S , \, N} f''(\nu)^p f'''(\nu)^q \equiv f(\nu)^S f'(\nu)^{-S+N}
 f''(\nu)^p f'''(\nu)^q  = \nu^4  
\end{equation}
assuming $p, q \geq 0$ (this describes all possible contribution terms) and,
without loss of generality, $f(\nu)$ a power law, we have solutions given by
\begin{equation}
\label{generalized ansatz}
f(\nu) \sim \nu^{(4 + N + 2p + 3q - S)/(N + p + q)}.
\end{equation}
with the added caveat that for $q \geq 0, f'''(\nu) \neq 0$.

All $\nu$-independent contributions to the corrected Friedmann equation can be
written as this type of differential equation and thus can only allow
solutions given by Eq.~(\ref{generalized ansatz}). In fact, as hinted at
previously, the general form includes expansions in higher positive and
negative powers of $b$ in Eq.~(\ref{general h}) because the higher powers
create more terms, but never qualitatively different results. For
instance, higher positive powers in $b$ are due to the power expansion of
$\sin b$ which will have the same $\nu$-dependence, that is $H_{2k+1} \sim H_5
\sim H_3 \sim H_1$ for non-negative integer $k$. Higher negative powers come
from expansions in the factor ordering which leads to varying powers of
$f'(\nu) / f(\nu)$, but this is included in the analysis above with an
appropriate choice of $S$ and $N$. What then is the lowest possible exponent
for $\nu$? For $f(\nu) \sim \nu^x$ where $x \leq 1$ we have:
\begin{equation}
  4+N+2p+3q - S \leq N + p + q \qquad \mbox{or equivalently}\qquad S \geq 4 + p + 2q
\end{equation}
However, as discussed above, the $S=4$ case does not allow $b$-independence in
the corrected Friedmann equation. Additionally, $p\geq 0$ and $q \geq 0$ since
they are formed from our higher derivative terms which always lead to positive
powers. Thus, we see that all possible cases that could lead to a term
independent of $\nu$ and $b$ in the effective Friedmann equation require
$f(\nu) \sim \nu^x$ with $x \geq 5/4$ which is incompatible with our
expectation of $f(\nu)$ within our framework for an isotropic, homogeneous
universe with a free, massless scalar field.

\section{Examples of truncated solutions}

We did not find a term independent of $\nu$ and $b$ even in the very
generalized case of factor orderings, which would have allowed us to provide
an explicit example for the effects suggested in
\cite{EffRecollapse}. However, with generic factor orderings considered, the
large-volume regime in quantum cosmology turns out to be much more subtle than
it appears in models with fixed (and simple) ordering. In particular, the
semiclassical expansion by powers of $\hbar$ leads to terms not analytic in
the curvature or Hubble parameter, so that the low-curvature regime is not
obvious. While we were able to provide (non-affirmative) indications regarding
the terms suggested in \cite{EffRecollapse}, the non-analytic nature of our
equations makes it difficult to obtain rigid statements about the possibility
or impossibility of a recollapse. In order to show some of the possible
behavior in the approach to low curvature, we now provide numerical solutions
of our previous equations with specific choices of coefficients of
non-analytic terms. Formally, these equations with a truncated $1/b$-expansion
allow possible recollapses, but near the recollapse with small $b$ any such
truncation breaks down. Only a non-truncated effective Hamiltonian could
provide a firm answer; however, without expanding the square root it is more
difficult to parameterize generic ordering ambiguities.

To provide an example, we will include only $H_1$ and $H_{-1}$ terms and
continue to focus on the regime of large volume and small curvature.
Our expanded Hamiltonian from Eq.~(\ref{general h}) then becomes
\begin{equation}
\label{general hneg1} H = H_1 b + H_{-1} b^{-1}\,,
\end{equation}
encapsulating only inverse-triad corrections and factor ordering, but no
holonomy modifications. (Note that non-analytic terms are mainly due to taking
a square root of a reordered Hamiltonian; they do not require the use of
holonomies and can therefore be realized in a Wheeler--DeWitt quantization.)
The quantum Hamiltonian to second order in moments is
\begin{equation}
\label{quantum hneg1} H_Q  =  H_1 b + H_{-1} b^{-1}  + \frac{1}{2} G^{\nu \nu}
( H''_1 b + H''_{-1} b^{-1} ) + G^{b \nu} ( H'_1 - H'_{-1} b^{-2})
+ G^{bb} H_{-1} b^{-3} \,.
\end{equation}
We solve for the full system of equations of motion by Poisson brackets of our
variables with the quantum Hamiltonian, where the dot represents
derivatives with respect to $\phi$ as before:
\begin{eqnarray}
\nonumber \dot{\nu} & = &  \frac{8 \pi G \gamma}{3} \left(
  \vphantom{\frac{G^{v v}}{2}} - H_1 + H_{-1} b^{-2} + 3 H_{-1} G^{b b} b^{-4}
  + \right. \\ 
& & \left.  \frac{G^{\nu \nu} }{2} (-H''_1 + H''_{-1} b^{-2} ) + 2 G^{b
    \nu} H'_{-1} b^{-3} \right) \\ 
\nonumber \dot{b} & = &    \frac{8 \pi G \gamma}{3} \left( H'_1 b +
  H'_{-1} b^{-1} + \frac{ G^{\nu \nu} }{2} (H'''_1 b + H'''_{-1}
  b^{-1} ) + \right. \\ 
& & \left. G^{b \nu} ( H''_1 - H''_{-1} b^{-2} ) + G^{bb} H'_{-1}
  b^{-3} \right) \\ 
 \dot{G}^{bb} & = &  \frac{8 \pi G \gamma}{3} \left( 2 G^{b \nu} (H''_1 b +
   H''_{-1} b^{-1} ) + 2 G^{bb} (H'_1 - H'_{-1} b^{-2}) \right) \\ 
 \dot{G}^{\nu \nu} & = &  \frac{8 \pi G \gamma}{3} \left( -4 G^{b \nu} H_{-1}
   b^{-3} - 2 G^{\nu \nu} (H'_1 - H'_{-1} b^{-1} ) \right) \\ 
 \dot{G}^{b \nu} & = &  \frac{8 \pi G \gamma}{3} \left( -2 G^{bb} H_{-1}
   b^{-3} + G^{\nu \nu} (H''_1 b + H''_{-1} b^{-1} ) \right)\,. 
\end{eqnarray}

As a further simplification, we truncate higher derivatives of $H_n$ as those
that correspond to increasing inverse powers of $\nu$. Keeping $H'_1 \sim
f'(\nu) \sim 1 +O(\nu^{-2})$ and $H_{-1} \sim \pm f'
(\nu)^2 \nu^{-1} $ to first order we are left with
\begin{eqnarray}
\label{bneg} \dot{b} & = &  \frac{8 \pi G \gamma}{3} b \\
\label{gbbneg} \dot{G}^{b b} & = &  \frac{8 \pi G \gamma}{3} 2 G^{b b} \\
\dot{G}^{b \nu} & = &  \frac{8 \pi G \gamma}{3} \left( \frac{ \mp 2 G^{bb}}{
    \nu b^{3}} \right) \\ 
\dot{G}^{\nu \nu} & = &  \frac{8 \pi G \gamma}{3} \left( \frac{ \mp 4 G^{b
      \nu}}{\nu b^{3}} - 2 G^{\nu \nu} \right) \\ 
\label{nunegend} \dot{\nu} & = &  \frac{8 \pi G \gamma}{3} \left( -\nu \pm \frac{1}{\nu b^{2}}
  \pm \frac{3 G^{b b}}{\nu b^{4} } \right) 
\end{eqnarray}
where the $\pm$ comes from $H_{-1}$ and the particular choice in factor
ordering. Our previous example had a negative sign, but in our general
discussion the sign --- or even numerical factors --- did not play a role.

We can solve this system of equations analytically. Equations~(\ref{bneg}) and
(\ref{gbbneg}) are independent of the other variables, which greatly simplify
the task. We easily obtain
\begin{equation} \label{bneg1} 
b  =  C_1 e^{\phi} \quad,\quad G^{bb}  =  C_2 e^{2 \phi}
\end{equation}
and with this continue to solve for
\begin{eqnarray}
\nonumber G^{\nu \nu}_\pm & = &  C_5 e^{-2 \phi} \mp \frac{4 e^{-2 \phi}}{C_1^2 (C_1^2 + 3 C_2)} \left(C_1 C_4 \sqrt{ C_1^4 C_3 \pm 2( C_1^2 + 3 C_2) \phi } \, \, \, \pm \right. \\
\label{gnunu+} & & \left. \frac{ C_2 ( C_1^4 C_3 \pm 2( C_1^2 + 3 C_2) \phi)} {C_1^2 + 3 C_2} \right) \\
G^{b \nu} & = & -\left( \frac{2 C_2 \sqrt{ \pm 2(C_1^2 + 3 C_2) \phi + C_1^4 C_3} }{
  C_1 (C_1^2 + 3 C_2) } \right) + C_4 \\
\label{nuneg1} \nu & = & \frac{e^{-\phi} \sqrt{\pm 2( C_1^2 + 3 C_2) \phi + C_1^4 C_3}   }{C_1^2}
\end{eqnarray}
where our coefficients can be chosen to meet our initial conditions
accordingly (and to respect the uncertainty relation for moments). Note that
depending on the sign of $H_{-1}$ we get $\pm$ in the equations for
$\nu(\phi)$ and $G^{b \nu}(\phi)$ as well as slightly different entries for
$G^{\nu \nu}(\phi)$ labeled by the $\pm$ subscript. Looking at
Eq.~(\ref{nuneg1}), we see that there is a possibility of negative $\phi$
within the squareroot, an indication that there could be turnover in the
volume as $\phi$ shrinks or grows (depending on the sign of the factor ordering
correction of $H_{-1}$). We also see that the constants under the square root
will play an important role, with $C_1 = b_0, C_2 = G^{bb}_0,$ and $C_3 =
\nu_0$.

\begin{figure}[tb]
\begin{center}
 \vspace{-2mm}
\includegraphics[width=0.8\textwidth]{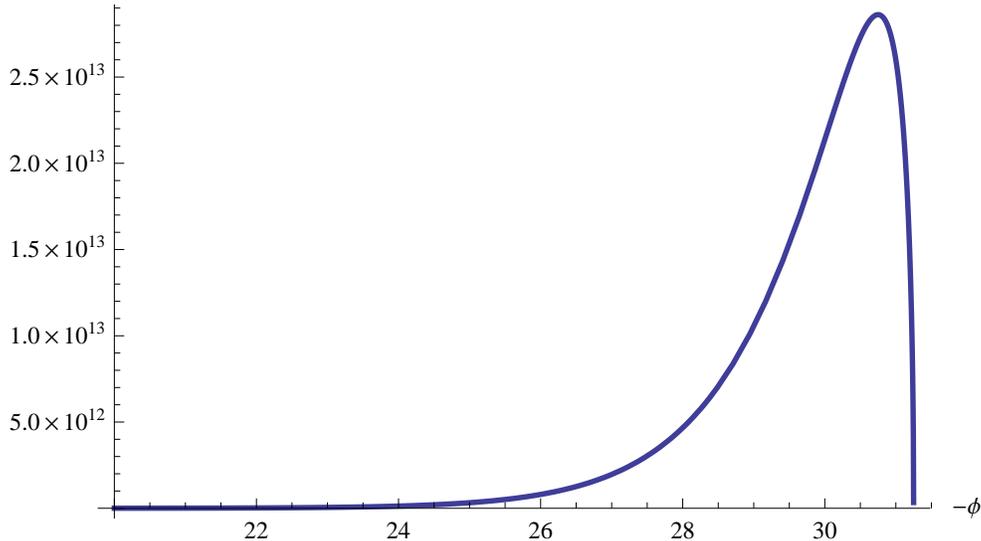}\vspace{-2mm}
\caption{Effective equation solution for the expectation value of volume $ \nu
  (\phi)$ is plotted with a $- \phi$ axis. The volume is exponentially
  increasing with $-\phi$
  over a large range of values until it begins a sharp
  descent when the square-root term of Eqn.~(\ref{nuneg1}) approaches zero at $
  - \phi \approx 30.75$.  
  \label{nucollapseHneg1}}
\end{center}
\vspace{-2mm}
\end{figure}

\begin{figure}[tb]
\begin{center}
 \vspace{-2mm}
\includegraphics[width=0.8\textwidth]{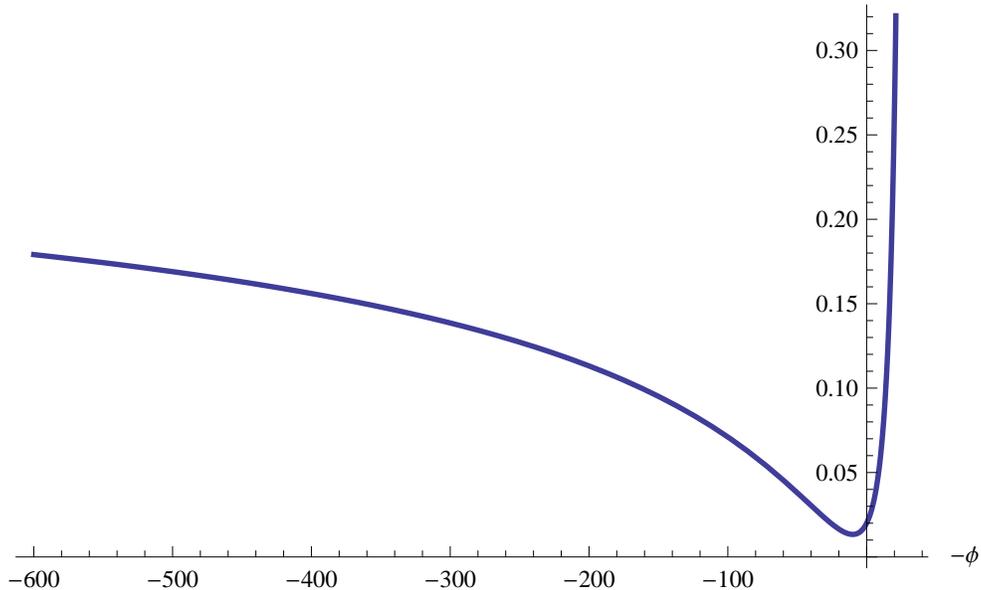}\vspace{-2mm}
\caption{Relative fluctuation of volume $ G^{\nu \nu} (\phi) / \nu(\phi)^2$
  is plotted with a $- \phi$ axis. It is well behaved over a large range of
  values but does increase sharply at $- \phi > 30.75$ and is increasingly
  larger for $ - \phi < -20$  where $\nu(\phi)$ becomes small.  
 \label{gvvRelHneg1}}
\end{center}
\vspace{-2mm}
\end{figure}

\begin{figure}[tb]
\begin{center}
 \vspace{-2mm}
\includegraphics[width=0.8\textwidth]{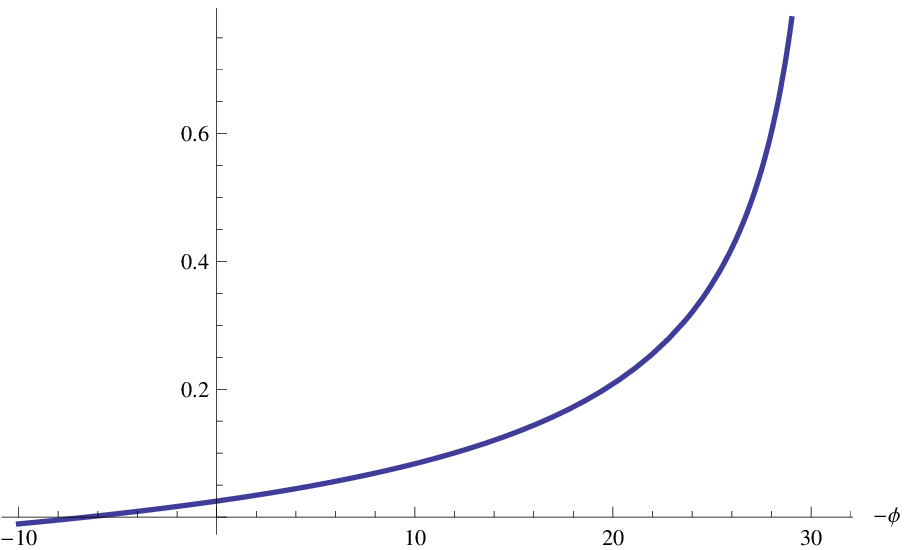}\vspace{-2mm}
\caption{Relative covariance fluctuation of the volume and curvature
  expectation values $ G^{b \nu} (\phi) / (b (\phi) \nu (\phi) )$ is plotted
  with a $- \phi$ axis. It is reasonably well behaved over a large range of
  values, though it increases unboundedly at large $- \phi$, which is
  inconsistent with our semiclassical approximations in that region. 
 \label{gbvRelHneg1}}
\end{center}
\vspace{-2mm}
\end{figure}

Let us then explore this system numerically for $H_{-1}$ positive, $\nu_0 =
10, b_0 = 1, G^{bb}_0 = 0.2, G^{\nu \nu}_0 = 2, G^{b \nu}_0 = 0.25$, noting that
our initial values satisfy the uncertainty relation (where we set $\hbar =1$
for our numerics)
\begin{equation}
\label{uncneg1}
G^{\nu \nu}G^{bb} - (G^{b \nu})^2 \geq \frac{\hbar^2}{4}.
\end{equation}
Using our truncated equations of motion, Eqs.~(\ref{bneg}) --
(\ref{nunegend}), we can easily verify that the uncertainty relation is
preserved, that is
\begin{equation}
\frac{{\rm d}}{{\rm d} \phi} \left( G^{\nu \nu}G^{bb} - (G^{b \nu})^2 \right) = 0
\end{equation}
and so will remain satisifed for all $\phi$. In Figure~\ref{nucollapseHneg1},
we see that our values do indeed lead to a collapse at late negative time
$\phi$ (plotted with $\phi$ increasingly negative to the right for
convenience, as we will do with all of this section's numerics). However, for
this plot to be meaningful, we must verify the consistency of our assumptions
within this region. In Figure~\ref{gvvRelHneg1} we see that the relative
volume fluctuation $ G^{\nu \nu} (\phi) / \nu^2 (\phi)$ is well behaved for
most of $\phi$, only increasing to the upper limits of our assumption of small
relative fluctuations when $\nu(\phi)$ becomes exponentially small for $ -
\phi < -20$. In fact, from our analytic solutions given by
Eqs.~(\ref{gnunu+}) and (\ref{nuneg1}), we see that the relative volume
fluctuations $G^{\nu \nu} / \nu^2$ asymptotically approaches a local maximum
for $ - \phi \rightarrow - \infty$:
\begin{equation}
\lim_{\phi \rightarrow \infty} \frac{G^{\nu \nu} (\phi)}{ \nu^2 (\phi)} = \frac{ 4 C_1^2 C_2}{(C_1^2 + 3C_2)^2} = \frac{ 4 b_0^2 G^{b b}_0 }{ (b_0^2 + 3 G^{b b}_0)^2}
\end{equation}
which is $0.3125$ for the numerical example used for our graphs (however,
graphing the large $\phi$ regime numerically would require a very large number
of integration steps). Near the turning point of $\nu$ at $-\phi > 30$, the
relative volume fluctuation begins to increase, quickly surpassing our
assumption of small relative values. From Eqs.~(\ref{bneg1}) we see that
relative curvature fluctuations $G^{bb} (\phi) / b^2 (\phi)$ will remain
constant according to where we set them by our constants $b_0$ and
$G^{bb}_0$.

In Figure~\ref{gbvRelHneg1} we see that the relative covariance $G^{b \nu} / (
b(\phi) \nu(\phi) )$ also surpasses our assumption of relatively small moments
as it approaches the turning point of $\nu$, becoming unboundedly large as
$\nu$ collapses at large $- \phi > 30 $. In addition, higher inverse powers of
$b$ contributing to the Hamiltonian with generic factor ordering will become
important near the potential recollapse. Thus, while we see collapse in
Figure~\ref{nucollapseHneg1}, it coincides with a breakdown of our assumptions
in the regime where $b$ becomes exponentially small: a regime where higher
inverse powers of curvature can become important, even when paired with
inverse-volume terms. (We can also explore the opposite sign choice for
$H_{-1}$ negative which is what our example from Subsection~\ref{factor
  ordering} had. In that case, the squareroot in $\nu(\phi)$ of
Eq.~(\ref{nucollapseHneg1}) goes to zero as $\phi \rightarrow
\infty$. However, rather than insinuating a recollapse, this is the regime
where $\nu(\phi) \rightarrow 0$ and $b(\phi)$ gets exponentially large,
indicating a singularity. Regardless of sign choice, in this regime we would
require a higher-curvature expansion by positive powers of $b$ for holonomy
(or other) corrections and quantum back-reaction before any physical meaning
can be drawn from such analysis.)

\section{Conclusion}

This article discusses the influence of factor-ordering choices on effective
equations in quantum cosmology, with possible modifications from loop
quantization. As we have emphasized, the quantization of the Hamiltonian
constraint is far from being unique even in the most reduced models. While
deparameterized models may offer simple quantization choices, they are not the
most general or natural ones. Quantum corrections and an analysis of
semiclassical physics must therefore take these ambiguities into account.

We have implemented such an analysis to study the question of whether quantum
effects in long-term semiclassical evolution could lead to significant
departures from classical behavior, for instance a recollapse of spatially
flat isotropic models. Since the model we analyzed allows a quantization free
of quantum back-reaction, low-curvature effects could only come from
factor-ordering corrections. We developed a suitable parameterization of
factor-ordering ambiguities in effective equations, paying special attention
to deparameterization choices. The generality of effective equations indeed
allows us to draw conclusions, indicating that too-drastic effects do not
occur. Our results therefore do not lead to concrete reasons for unexpected
late-time effects in semiclassical (loop) quantum cosmology. But they provide
strong caution against too-quick positive assurances of a straightforward
classical limit based on the analysis of a few simple models and orderings.

In particular, it does not appear possible to have quantum corrections of the
modified Friedmann equation independent of the energy density in this
formulation. Thus, quantum collapse at low energy densities or large scales
does not seem feasible by this mechanism given our corrections and
assumptions. The only possibility, according to our analysis, would be to have
inverse-triad corrections with a function $f(\nu)$ increasing more strongly
than linearly, but this is not acceptable by the classical limit of the
Hamiltonian constraint.

Furthermore, we explored the effects of factor ordering choices at first
semiclassical order. A general tractable parameterization of quantization
ambiguities in this regime surprisingly turns out to require a second
expansion by the inverse Hubble parameter. Although the solutions to these
effective equations hint at the possibility of a recollapse, the expansion by
the inverse Hubble parameter breaks down before a recollapse could be reached.
Interestingly, also some relative fluctuations grow too large to satisfy our
approximations for all $\phi$ of interest. Extending such an analysis to
higher quantum moments could be fruitful; for instance, unlimited
semiclassical evolution may be possible within our expansion scheme if there
are corrections to $b(\phi)$ such that it does not asymptotically approach
zero, thus preventing factor ordering terms $1/b$ to approach infinity. For
now, however, the semiclassical behavior of generic quantum cosmology remains
incompletely understood.

\section*{Acknowledgements}

We are grateful to Yongge Ma for several discussions and for hospitality at
Beijing Normal University during an early stage of this project.  This
research was supported in part by the NSF East Asia and Pacific Summer
Institute Fellowship to DS, by NSF grants PHY-0748336 and PHY-1307408, and by
Perimeter Institute for Theoretical Physics during a visit of MB. Research at
Perimeter Institute is supported by the Government of Canada through Industry
Canada and by the Province of Ontario through the Ministry of Economic
Development \& Innovation.


\begin{thebibliography}{10}

\bibitem{BouncePert}
M.\ Bojowald,
\newblock Large scale effective theory for cosmological bounces,
\newblock {\em Phys.\ Rev.\ D} 75 (2007) 081301(R), [gr-qc/0608100]

\bibitem{EffRecollapse}
Y.\ Ding, Y.\ Ma, and J.\ Yang,
\newblock Effective Scenario of Loop Quantum Cosmology,
\newblock {\em Phys.\ Rev.\ Lett.} 102 (2009) 051301, [arXiv:0808.0990]

\bibitem{Taveras}
V.\ Taveras,
\newblock Corrections to the Friedmann Equations from LQG for a Universe with a
  Free Scalar Field,
\newblock {\em Phys.\ Rev.\ D} 78 (2008) 064072, [arXiv:0807.3325]

\bibitem{Bohr}
A.\ Ashtekar, M.\ Bojowald, and J.\ Lewandowski,
\newblock Mathematical structure of loop quantum cosmology,
\newblock {\em Adv.\ Theor.\ Math.\ Phys.} 7 (2003) 233--268, [gr-qc/0304074]

\bibitem{QCReview}
D.~L.\ Wiltshire,
\newblock An introduction to quantum cosmology,
\newblock In B.\ Robson, N.\ Visvanathan, and W.~S.\ Woolcock, editors, {\em
  Cosmology: The Physics of the Universe}, pages 473--531. World Scientific,
  Singapore, 1996, [gr-qc/0101003]

\bibitem{LivRev}
M.\ Bojowald,
\newblock Loop Quantum Cosmology,
\newblock {\em Living Rev.\ Relativity} 11 (2008) 4, [gr-qc/0601085],
\newblock {\tt http://www.livingreviews.org/lrr-2008-4}

\bibitem{Springer}
M.\ Bojowald,
\newblock {\em Quantum Cosmology: A Fundamental Theory of the Universe},
\newblock Springer, New York, 2011

\bibitem{Rov}
C.\ Rovelli,
\newblock {\em Quantum Gravity},
\newblock Cambridge University Press, Cambridge, UK, 2004

\bibitem{ThomasRev}
T.\ Thiemann,
\newblock {\em Introduction to Modern Canonical Quantum General Relativity},
\newblock Cambridge University Press, Cambridge, UK, 2007, [gr-qc/0110034]

\bibitem{ALRev}
A.\ Ashtekar and J.\ Lewandowski,
\newblock Background independent quantum gravity: A status report,
\newblock {\em Class.\ Quantum Grav.} 21 (2004) R53--R152, [gr-qc/0404018]

\bibitem{LoopRep}
C.\ Rovelli and L.\ Smolin,
\newblock Loop Space Representation of Quantum General Relativity,
\newblock {\em Nucl.\ Phys.\ B} 331 (1990) 80--152

\bibitem{SymmRed}
M.\ Bojowald and H.~A.\ Kastrup,
\newblock Symmetry Reduction for Quantized Diffeomorphism Invariant Theories of
  Connections,
\newblock {\em Class.\ Quantum Grav.} 17 (2000) 3009--3043, [hep-th/9907042]

\bibitem{NonAb}
M.\ Bojowald,
\newblock Mathematical structure of loop quantum cosmology: Homogeneous models,
  [arXiv:1206.6088]

\bibitem{DegFull}
M.\ Bojowald,
\newblock Degenerate Configurations, Singularities and the Non-Abelian Nature
  of Loop Quantum Gravity,
\newblock {\em Class.\ Quantum Grav.} 23 (2006) 987--1008, [gr-qc/0508118]

\bibitem{Karpacz}
M.\ Bojowald and A.\ Skirzewski,
\newblock Quantum Gravity and Higher Curvature Actions,
\newblock {\em Int.\ J.\ Geom.\ Meth.\ Mod.\ Phys.} 4 (2007) 25--52,
  [hep-th/0606232]

\bibitem{HigherTime}
M.\ Bojowald, S.\ Brahma, and E.\ Nelson,
\newblock Higher time derivatives in effective equations of canonical quantum
  systems,
\newblock {\em Phys.\ Rev.\ D} 86 (2012) 105004, [arXiv:1208.1242]

\bibitem{AmbigConstr}
K.\ Vandersloot,
\newblock On the Hamiltonian Constraint of Loop Quantum Cosmology,
\newblock {\em Phys.\ Rev.\ D} 71 (2005) 103506, [gr-qc/0502082]

\bibitem{RSLoopDual}
P.\ Singh,
\newblock Loop cosmological dynamics and dualities with Randall-Sundrum
  braneworlds,
\newblock {\em Phys.\ Rev.\ D} 73 (2006) 063508, [gr-qc/0603043]

\bibitem{APSII}
A.\ Ashtekar, T.\ Pawlowski, and P.\ Singh,
\newblock Quantum Nature of the Big Bang: Improved dynamics,
\newblock {\em Phys.\ Rev.\ D} 74 (2006) 084003, [gr-qc/0607039]

\bibitem{BounceSqueezed}
M.\ Bojowald,
\newblock How quantum is the big bang?,
\newblock {\em Phys.\ Rev.\ Lett.} 100 (2008) 221301, [arXiv:0805.1192]

\bibitem{cosmoIV}
M.\ Bojowald,
\newblock Loop Quantum Cosmology IV: Discrete Time Evolution,
\newblock {\em Class.\ Quantum Grav.} 18 (2001) 1071--1088, [gr-qc/0008053]

\bibitem{IsoCosmo}
M.\ Bojowald,
\newblock Isotropic Loop Quantum Cosmology,
\newblock {\em Class.\ Quantum Grav.} 19 (2002) 2717--2741, [gr-qc/0202077]

\bibitem{SemiClass}
M.\ Bojowald,
\newblock The Semiclassical Limit of Loop Quantum Cosmology,
\newblock {\em Class.\ Quantum Grav.} 18 (2001) L109--L116, [gr-qc/0105113]

\bibitem{RegularizationFRW}
J.\ Haro and E.\ Elizalde,
\newblock Effective gravity formulation that avoids singularities in quantum
  FRW cosmologies, [arXiv:0901.2861]

\bibitem{RegularizationLQC}
R.\ Helling,
\newblock Higher curvature counter terms cause the bounce in loop cosmology,
  [arXiv:0912.3011]

\bibitem{EffAc}
M.\ Bojowald and A.\ Skirzewski,
\newblock Effective Equations of Motion for Quantum Systems,
\newblock {\em Rev.\ Math.\ Phys.} 18 (2006) 713--745, [math-ph/0511043]

\bibitem{AltCollapse}
J.\ Yang, Y.\ Ding and Y.\ Ma, Alternative quantization of the Hamiltonian
in loop quantum cosmology, {\em Phys.\ Lett.\ B} 682 (2009) 1--7,
[arXiv:0904.4379]

\bibitem{FuncCollapse1}
L.\ Qin and Y.\ Ma, Coherent state functional integrals in quantum
cosmology, {\em Phys.\ Rev.\ D} 85 (2012) 063515, [arXiv:1110.5480]

\bibitem{FuncCollapse2}
L.\ Qin and Y.\ Ma, Coherent state functional Integral in Loop Quantum
Cosmology: Alternative Dynamics, {\em Mod.\ Phys.\ Lett. A} 27 (2012) 1250078,
[arXiv:1206.1128]

\bibitem{BounceCohStates}
M.\ Bojowald,
\newblock Dynamical coherent states and physical solutions of quantum
  cosmological bounces,
\newblock {\em Phys.\ Rev.\ D} 75 (2007) 123512, [gr-qc/0703144]

\bibitem{AshVarReell}
J.~F.\ Barbero~G.,
\newblock Real Ashtekar Variables for Lorentzian Signature Space-Times,
\newblock {\em Phys.\ Rev.\ D} 51 (1995) 5507--5510, [gr-qc/9410014]

\bibitem{Immirzi}
G.\ Immirzi,
\newblock Real and Complex Connections for Canonical Gravity,
\newblock {\em Class.\ Quantum Grav.} 14 (1997) L177--L181

\bibitem{InhomLattice}
M.\ Bojowald,
\newblock Loop quantum cosmology and inhomogeneities,
\newblock {\em Gen.\ Rel.\ Grav.} 38 (2006) 1771--1795, [gr-qc/0609034]

\bibitem{QSDV}
T.\ Thiemann,
\newblock {QSD V}: Quantum Gravity as the Natural Regulator of Matter Quantum
  Field Theories,
\newblock {\em Class.\ Quantum Grav.} 15 (1998) 1281--1314, [gr-qc/9705019]

\bibitem{InvScale}
M.\ Bojowald,
\newblock Inverse Scale Factor in Isotropic Quantum Geometry,
\newblock {\em Phys.\ Rev.\ D} 64 (2001) 084018, [gr-qc/0105067]

\bibitem{LoopMuk}
M.\ Bojowald and G.\ Calcagni,
\newblock Inflationary observables in loop quantum cosmology,
\newblock {\em JCAP} 1103 (2011) 032, [arXiv:1011.2779]

\bibitem{EffCons}
M.\ Bojowald, B.\ Sandh\"ofer, A.\ Skirzewski, and A.\ Tsobanjan,
\newblock Effective constraints for quantum systems,
\newblock {\em Rev.\ Math.\ Phys.} 21 (2009) 111--154, [arXiv:0804.3365]

\bibitem{HighDens}
M.\ Bojowald, D.\ Mulryne, W.\ Nelson, and R.\ Tavakol,
\newblock The high-density regime of kinetic-dominated loop quantum cosmology,
\newblock {\em Phys.\ Rev.\ D} 82 (2010) 124055, [arXiv:1004.3979]

\bibitem{HigherMoments}
M.\ Bojowald, D.\ Brizuela, H.~H.\ Hernandez, M.~J.\ Koop, and H.~A.\
  Morales-T\'ecotl,
\newblock High-order quantum back-reaction and quantum cosmology with a
  positive cosmological constant,
\newblock {\em Phys.\ Rev.\ D} 84 (2011) 043514, [arXiv:1011.3022]

\bibitem{QSDI}
T.\ Thiemann,
\newblock Quantum Spin Dynamics {(QSD)},
\newblock {\em Class.\ Quantum Grav.} 15 (1998) 839--873, [gr-qc/9606089]

\bibitem{Ambig}
M.\ Bojowald,
\newblock Quantization ambiguities in isotropic quantum geometry,
\newblock {\em Class.\ Quantum Grav.} 19 (2002) 5113--5130, [gr-qc/0206053]

\bibitem{ICGC}
M.\ Bojowald,
\newblock Loop Quantum Cosmology: Recent Progress,
  {\em Pramana} 63 (2004) 765--776, [gr-qc/0402053]

\bibitem{Action}
M.\ Bojowald and G.~M.\ Paily,
\newblock Deformed General Relativity and Effective Actions from Loop Quantum
  Gravity,
\newblock {\em Phys.\ Rev.\ D} 86 (2012) 104018, [arXiv:1112.1899]

\bibitem{SigChange}
J.\ Mielczarek,
\newblock Signature change in loop quantum cosmology, [arXiv:1207.4657]

\bibitem{QuantumBounce}
M.\ Bojowald,
\newblock Quantum nature of cosmological bounces,
\newblock {\em Gen.\ Rel.\ Grav.} 40 (2008) 2659--2683, [arXiv:0801.4001]

\bibitem{QCEffCons}
X.\ Wu and Y.\ Ma,
\newblock Effective Theories of Quantum Cosmology, [arXiv:1212.5874]

\bibitem{ReducedKasner}
P.\ Malkiewicz,
\newblock Reduced phase space approach to Kasner universe and the problem of
  time in quantum theory,
\newblock {\em Class.\ Quantum Grav.} 29 (2012) 075008, [arXiv:1105.6030]

\bibitem{Electric}
S.\ Alexander, M.\ Bojowald, A.\ Marciano, and D.\ Simpson,
\newblock Electric time in quantum cosmology,
\newblock {\em Class.\ Quantum Grav.} (2013) to appear, [arXiv:1212.2204]

\bibitem{EffTime}
M.\ Bojowald, P.~A.\ H\"ohn, and A.\ Tsobanjan,
\newblock An effective approach to the problem of time,
\newblock {\em Class.\ Quantum Grav.} 28 (2011) 035006, [arXiv:1009.5953]

\bibitem{EffTimeLong}
M.\ Bojowald, P.~A.\ H\"ohn, and A.\ Tsobanjan,
\newblock An effective approach to the problem of time: general features and
  examples,
\newblock {\em Phys.\ Rev.\ D} 83 (2011) 125023, [arXiv:1011.3040]

\bibitem{EffTimeCosmo}
P.~A.\ H\"ohn, E.\ Kubalova, and A.\ Tsobanjan,
\newblock Effective relational dynamics of a nonintegrable cosmological model,
\newblock {\em Phys.\ Rev.\ D} 86 (2012) 065014, [arXiv:1111.5193]

\end{thebibliography}

\end{document}